\documentclass[twocolumn,pra,aps,superscriptaddress]{revtex4}
\usepackage{graphicx}
\usepackage{bm}

\def\kvec{\bm{k}}

\def\ak{a_{\bm k}}
\def\akplus{a_{\bm k}^\dagger}
\def\omk{\omega_{\bm k}}

\def\dt{\!{\rm d}}

\def\duzomniejsze{<\kern-1.5mm<}
\def\duzowieksze{>\kern-.7mm>}

\def\textbf#1{{\bf #1}}
\def\beq{\begin{equation}}
\def\eeq{\end{equation}}
\def\be{\begin{equation}}
\def\ee{\end{equation}}
\def\ben{\begin{eqnarray}}
\def\een{\end{eqnarray}}
\def\beqa{\begin{eqnarray}}
\def\eeqa{\end{eqnarray}}
\def\eea{\end{array}}
\def\bea{\begin{array}}
\newcommand{\bei}{\begin{itemize}}
\newcommand{\eei}{\end{itemize}}
\newcommand{\bee}{\begin{enumerate}}
\newcommand{\eee}{\end{enumerate}}

\def\>{\rangle}
\def\<{\langle}

\begin{document}
\title{Optimal strategy for a single-qubit gate and trade-off between
opposite types of decoherence}

\author{Robert Alicki}
\author{Micha\l{} Horodecki}
\affiliation{Institute of Theoretical Physics and Astrophysics,
University of Gda\'{n}sk, 80-952 Gda\'{n}sk, Poland}
\author{Pawe\l{} Horodecki}
\affiliation{Faculty of Applied Physics and Mathematics, Technical
University of Gda\'nsk, 80-952 Poland}
\author{Ryszard Horodecki}
\affiliation{Institute of Theoretical Physics and Astrophysics,
University of Gda\'{n}sk, 80-952 Gda\'{n}sk, Poland}
\author{Lucjan Jacak}
\affiliation{Institute of Physics, Wroc\l{}aw University of
Technology, 50-370 Wroc{\l}aw, Poland}
\author{Pawe\l{} Machnikowski}
\affiliation{Institute of Physics, Wroc\l{}aw University of
Technology, 50-370 Wroc{\l}aw, Poland}
\affiliation{Institut f\"ur
Festk\"orpertheorie, Westf\"alische Wilhelms-Universit\"at, 48149
M\"unster, Germany}

\begin{abstract}
We study reliable quantum information processing (QIP) under two
different types of environment. First type is Markovian
exponential decay, and the appropriate elementary strategy of
protection of qubit is to apply fast gates. The second one is
strongly non-Markovian and occurs solely during operations on the
qubit.  The best strategy is then to work with slow gates. If the
two types are both present, one has to optimize the speed of gate.
We  show that such a trade-off is present for a single-qubit
operation in a semiconductor quantum dot implementation of QIP,
where recombination of exciton (qubit) is Markovian, while phonon
dressing gives rise to the non-Markovian contribution.
\end{abstract}

\maketitle

Already at the early stage of quantum information theory,
the implementation of quantum computing was found to be extremely
challenging due to decoherence processes \cite{Unruh1995}.
It was claimed, nevertheless, that a quantum computer could
reliably perform quantum algorithms once the error per
operation is below some threshold level which is approximately
$\delta\approx 10^{-5}$ \cite{QEC}
(for some caveats coming from memory of environment see
\cite{alicki02a}).
In any case, minimizing the error per gate is a reasonable strategy.
Usually, one considers
the decoherence time of the system $\tau_{\mathrm{d}}$ and the time of
performing operation (quantum gate) $\tau_{\mathrm{g}}$, so that the error is given by
the ratio $\tilde{\delta}=\tau_{\mathrm{g}}/\tau_{\mathrm{d}}$
\cite{DiVincenzo95}.
This suggests to search for systems with $\tilde{\delta}$
as small as possible, and the obvious strategy to diminish the error is to
apply fast gates. One then tacitly assumes that the process of
decoherence is independent of running the gate,
which formally means that it is a Markovian process, where the error,
to first order, grows linearly in time.
Gate speed-up  may be achieved
by selecting materials to provide favorable spectrum characteristics
\cite{derinaldis02} or by applying
techniques reducing unwanted transitions
within the register space \cite{chen} as well
as outside this space (leakage) \cite{cancel}.

The assumption of Markovian character of noise is by no means
obvious and, indeed, it has recently been found \cite{alicki02a}
that due to non-Markovian effects it may prove to be completely
invalid. The notion of {\it minimal decoherence model} was
introduced where the error occurs {\it solely} during gate
operation, and it grows for fast gates as $\delta\sim
1/\tau_{\mathrm{g}}^2$ (for the spectral density of the reservoir
$\sim\omega^{3}$). An example of such a model is a degenerate
system interacting with a bosonic field at vacuum via dipole
interaction. This kind of error favors the counterintuitive
strategy of \textit{slowing down} the gate operation.

In real systems, this type of decoherence competes with the
Markovian damping, although this effect may be covered by other
errors (e.g. leakage). In this paper we point out that such a
competitive effect is essential e.g. for the solid-state qubit
implementation using excitonic (charge) states in quantum dots
(QDs) \cite{Rossi-SQUID2000}, with computational states defined by
the absence ($|0\rangle$) or presence ($|1\rangle$) of one exciton
in the ground state of the dot, operated by resonant coupling to
laser light. In such a system, Markovian decoherence (exponential
damping) is related to exciton recombination on 1 ns timescale,
while strongly non-Markovian effects result from lattice inertia.
The interplay of these two decoherence mechanisms, favoring
opposite strategies (fast vs. slow gates), leads to a kind of
trade-off, resulting in optimal speed of gate for most reliable
operation on the qubit.

To investigate the effect, we analyze decoherence in a general
spin-boson system \cite{vanHove,Alicki-spinbozon}, and find
decoherence to be strongly non-Markovian, fitting into ``minimal
decoherence model''. Assuming additional, Markovian damping, we
obtain the trade-off formula for the error caused by decoherence,
averaged over input states of the qubit \be \delta =
{\gamma_{\mathrm{nM}} \over \tau_{\mathrm{g}}^2} +
\gamma_{\mathrm{M}} \tau_{\mathrm{g}} \ee (actually, in the solid
state example, singularity of the first term is lifted by the
presence of upper cut-off). The constants $\gamma_{\mathrm{M,nM}}$
express the strength of the Markovian and non-Markovian
decoherence, respectively. We determine their values for the
exciton in a QD with typical parameters and show that the gate
duration leading to minimal overall error is of the order of $1$
ps.
Since level spacing in such a system allows even 100 fs gating
\cite{chen} the restrictions imposed by non-Markovian phonon
mechanisms are decisive for the gate speed. Our result has both
practical as well as  general implications for QIP: (i) it
suggests the proper direction in research towards semiconductor
implementation of QIP (ii) it provides a non-trivial dynamical
strategy to minimize decoherence in quantum systems. for a class
of reservoirs.


{\it Minimal decoherence in the spin-boson model.}
We consider a qubit described by means of the spin-boson model.
The Hamiltonian $H''$ of the system plus reservoir is
 \be \label{ham}
    H''= H_S^0+H_{S}(t) + H_{SR} + H_R,
 \ee
where $S$ is our qubit system, $R$ is (bosonic) reservoir.
$H_S^0=(1/2)\Omega \sigma_z$  is the self Hamiltonian of the
qubit, $H_S(t)=(1/2)\epsilon(t) (e^{i\Omega t} \sigma_+ +
e^{-i\Omega t}\sigma_-)$ is the  gate Hamiltonian (in rotating
wave approximation), $\epsilon(t)$ being the shape of the pulse.
$H_R=\sum \omk \ak \akplus $ is the self Hamiltonian of the
reservoir, where $\omk$ is the boson energy. Finally, $H_{SR}$ is
the interaction with reservoir
 \be
    H_{SR}= \sigma_z\otimes \left[ \sum f_{\kvec}^{*} \ak +
        \mathrm{H.c.}\right],
 \ee
 where $f_{\kvec}$ is the coupling constant for the mode $\kvec$.

In the interaction picture with respect to $H_S^0$ we get
$H'=\epsilon(t) \sigma_x + H_{SR} + H_R.$ Let us now represent
$H'$ in the basis of eigenstates of the total system (dressed
states). This is achieved by the unitary operation $U=
|0\>\<0|\otimes W-|1\>\<1| \otimes W^{\dag}$, where
$W=\exp[\sum(f^{*}_{\kvec}/\omega_{\kvec})a_{\kvec}-\mathrm{H.c}]$
and $|0\>, |1\>$ are the eigenvectors of $\sigma_z$.
In this basis the Hamiltonian $H=U^{\dagger}H'U$ up to a constant
has the form
 \ben
H & = & \frac{1}{2}\epsilon(t) \sigma_x \cos\left(2i
\sum_{\kvec}\frac{f_{\kvec}}{\omega_{\kvec}}a_{\kvec}^{\dag}
    +\mathrm{H.c.}\right)
\\ \nonumber
&&+\frac{1}{2}\epsilon(t) \sigma_y \sin\left(2i
\sum_{\kvec}\frac{f_{\kvec}}{\omega_{\kvec}}a_{\kvec}^{\dag}
    +\mathrm{H.c.}\right)
+H_R.
 \een
To the first order in $f_{\kvec}/\omega_{\kvec}$ we obtain
 \be
    H\simeq \epsilon (t)\sigma_x + \epsilon(t) \sigma_y \otimes
    \left(i
    \sum_{\kvec}\frac{f_{\kvec}}{\omega_{\kvec}}a^{\dag}+\mathrm{H.c.}
    \right) +H_R.
 \ee

We define the qubit in terms of the dressed states
\cite{Alicki-spinbozon}. With such a choice, the reservoir is
decoupled and the interaction term is present only during gate
operation. Thus our system may be reduced to the minimal
decoherence model: decoherence is not present at all, if the qubit
is not active.
Now we want to calculate the resulting error
and relate  it to the speed of gate. Since the regime is
strongly non-Markovian we solve the Master equation in the Born
approximation and compute the fidelity $F=1-\delta$ of a single-qubit
operation (see \cite{alicki02a} for details).


The error, averaged over the initial qubit states can be
represented as the overlap of two functions
 \be\label{delta}
    \delta= \int \frac{\dt \omega}{\omega^2} R(\omega)S(\omega).
 \ee
 Here $R(\omega)$ is the spectral density of the reservoir
 \be
    R(\omega)=[n_{\mathrm{B}}(\omega)+1]\sum_{\mathbf{k}}
    \left[ \delta (\omega_{\mathbf{k}}-\omega)
        + \delta (\omega_{\mathbf{k}} +\omega) \right] |f_{\kvec}|^2,
 \label{spectral}
 \ee
 where $n_{\mathrm{B}}(\omega)$ is the Bose-Einstein distribution.
The function $S(\omega)$ fully represents the spectral
characteristics of the system and is given by
\begin{eqnarray}
  S(\omega) &=&  \left[\<\psi|YY^\dagger|\psi\>- \<\psi|Y^\dagger|\psi\>
    \<\psi|Y|\psi\>\right]_{\mathrm{av}} \\ \nonumber
 &\approx& \frac{1}{3} (|F_-(\omega)|^2+|F_+(\omega)|^2),
\end{eqnarray}
where $[]_{\mathrm{av}}$ denotes averaging over the states $\psi$
and
\begin{eqnarray*}
   Y=i\big(F_+ |+\>\<-| + F_- |-\>\<+| \big),
  && \; |\pm\>=\frac{|0\>\pm|1\>}{\sqrt{2}}; \\
    F_{\pm}=\pm\int_{-\infty}^{+\infty}\dt u \,\, e^{\pm i\phi(u)}
    \epsilon(u) e^{i\omega u};
  && \; \phi(t)=\int_{-\infty}^t \dt u \epsilon(u).
\end{eqnarray*}

A complete minimization of $\delta$ would require full
optimization of the pulse shape. However, in order to demonstrate
the idea of the trade-off in simple terms, let us restrict the
discussion to qubit rotations performend by Gaussian pulses,
$\epsilon(t)= {\alpha / (\sqrt{2\pi} \tau_{\mathrm{g}})}
e^{-{1\over 2} \big( {t/\tau_{\mathrm{g}} }\big)^2}$. Here
$\tau_{\mathrm{g}} $ is the gate duration, while $\alpha$ is the
angle determining the gate, e.g. $\alpha={\pi\over 2}$ is
$\sqrt{\mathrm{NOT}}$, while $\alpha=\pi$ is $\sigma_x$ (bit
flip).

The functions $|F_\pm(\omega)|^2$ may be written as
\begin{equation}\label{Fpm}
|F_{\pm}(\omega)|^{2}\approx \alpha^{2} e^{-\tau_{g}^{2} \left(
\omega\pm \frac{\alpha}{\sqrt{2\pi}\tau_{g}} \right)^{2}}.
\end{equation}

\begin{figure}[tb]
\unitlength 1mm
\begin{center}
\begin{picture}(80,48)(0,5)
\put(0,0){\resizebox{80mm}{!}{\includegraphics{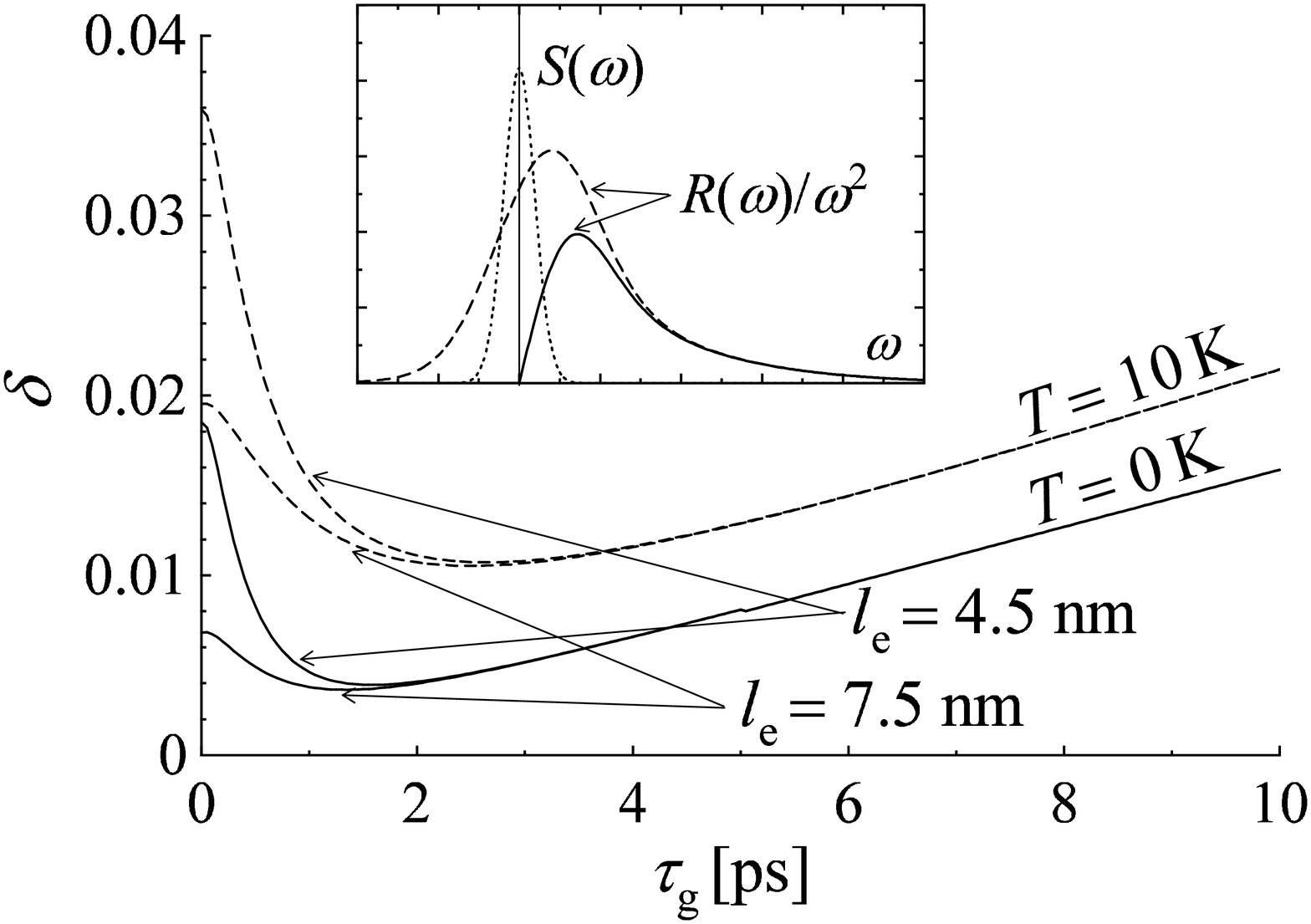}}}
\end{picture}
\end{center}
\caption{\label{fig:result}The total error for a $\alpha=\pi/2$
rotation on a QD qubit, for $T=0$ (solid lines) and $T=10$ K
(dashed lines), for two dot sizes ($l_{h}=0.8l_{e}$,
$l_{z}=0.2l_{e}$). The Markovian decoherence times are inferred
from the experimental data \protect\cite{borri01}. Inset: Spectral
density of the phonon reservoir $R(\omega)$ at these two
temperatures and the gate profile $S(\omega)$ for $\alpha=\pi/2$.}
\end{figure}

Although the original minimal decoherence model with its $\sim
\omega^{3}$ dependence appears in many physical situations, other
characteristics are obviously possible. In general, as may be seen
from (\ref{delta}), (\ref{spectral}) and (\ref{Fpm}), for a
spectral density $R(\omega)\sim \omega^{n}$ the error scales with
the gate duration as $\tau_{\mathrm{g}}^{-n+1}$ and
$\tau_{\mathrm{g}}^{-n+2}$ at low and high temperatures,
respectively. Therefore, for $n>2$ (typical e.g. for various types
of phonon reservoirs \cite{krummheuer02}) the error grows for
faster gates. Assuming the spectral density of the form
$R(\omega)=R_{0}\omega^3$ (see the QD example below), we obtain
 \be
    \delta_{\mathrm{nM}} =
    \frac{1}{3} \alpha^2 R_0 \tau_{\mathrm{g}}^{-2},\;\;\mbox{at}\;T=0
 \ee
This leading order formula holds for $\delta \duzomniejsze 1$.
Also, if we introduce the upper cut-off, the error will be finite
even for an infinitely fast gate \cite{jacak03b} (see Fig.
\ref{fig:result}).

{\it Trade-off between two types of decoherence}. As we have
shown,  in our model the error grows as the speed of gate
increases. This could result in obtaining arbitrarily low error by
choosing suitably low speed of gates. However, if the system is
also subject to other types of noise this becomes  impossible.
Indeed, assuming an additional contribution growing with rate
$\gamma_{\mathrm{M}}$, the total error per gate is
\begin{equation}
\delta=\frac{\gamma_{\mathrm{nM}}}{\tau_{\mathrm{g}}^{2}}
+\gamma_{\mathrm{M}}\tau_{\mathrm{g}}, \;\;
\gamma_{\mathrm{nM}}=\frac{1}{3}\alpha^{2}R_{0},\;\;
\gamma_{\mathrm{M}}=\frac{1}{\tau_{\mathrm{r}}},
\end{equation}
where $\tau_{\mathrm{r}}$ is the characteristic time of Markovian
decoherence (recombination time in the excitonic case).
As a result, the overall error is unavoidable and optimization is needed.
The above formulas lead to the optimal values of the form (for $T=0$)
\begin{equation}
\delta_{\mathrm{min}}
=\frac{3}{2} \left(
\frac{2\alpha^{2}R_{0}}{3\tau_{\mathrm{r}}^{2}} \right)^{1/3},\;\;
\mbox{for}\; \tau_{\mathrm{g}}
=\left( \frac{2}{3}\alpha^{2}R_{0}\tau_{\mathrm{r}} \right)^{1/3}.
\label{eq:minerror}
\end{equation}
We will see that the trade-off  is present  also for
finite temperatures.

{\it Example: exciton in a quantum dot (QD).} Let us now estimate
the error magnitude for the specific semiconductor QD qubit
implementation \cite{Rossi-SQUID2000}. The reservoir is then
constituted by phonons, the most important branch for our present
purpose being the longitudinal acoustical (LA) one, characterized
by linear dispersion, $\omega_k=ck$, coupled via deformation
potential (DP) \cite{mahan00}. In fact, the relatively simple
model (\ref{ham}) is very accurate for a description of the QD
system for $\sim 1$ ps timescales relevant here, as confirmed by
the excellent agreement between the theoretical calculations
\cite{vagov} and experimental results \cite{borri01}. This is due
to the fact that neither the high-frequency optical phonons nor
direct or phonon-induced leakage to higher states contribute
considerably to the decoherence. Also effects related to
piezoelectric coupling to LA and TA phonons may be neglected in
weakly piezoelectric systems (e.g. GaAs). Moreover, the qubit
control is all-optical, eliminating the need for additional
noise-inducing device structures. On the other hand, details of
the QD structure (shape, stress, composition) may lead only to
quantitative modifications of secondary importance.

The non-Markovian error has a dynamical origin and is related to a
which path trace left by exciting the phonon modes rather than to
an influence of noise. The spectral density is uniquely defined by
the lattice response characteristics: the mode-dependent coupling
strength and the density of states. Due to fundamental
restrictions (global charge neutrality, translational invariance),
the frequency dependence is always strongly super-ohmic
\cite{krummheuer02}. On the other hand, the approximate momentum
conservation holding for weakly confined carrier states leads to
exponential cut-off of carrier-phonon interaction at high
frequencies \cite{jacak}. In the specific case of DP coupling, the
coupling constants are \cite{mahan00}
 \be
    f_{\kvec} =\frac{1}{2} \sqrt{\frac{k}{2\varrho v c}} \left(
    \sigma_{\mathrm{e}} {\cal F}^{(\mathrm{e})}_{\kvec}
    -\sigma_{\mathrm{h}} {\cal F}^{(\mathrm{h})}_{\kvec} \right),
 \label{coupling}
 \ee
 where $v$ is the volume of the unit cell,
$\rho$ is the crystal density, $\sigma_{\mathrm{e,h}}$ are
deformation potential constants for the electron and the hole
(material parameters are taken as in \cite{krummheuer02}) and
${\cal F}^{\mathrm{(e,h)}}_{\kvec}$ are formfactors for the
corresponding wavefunctions (approximated by Gaussians, on the
grounds of numerical diagonalization in parabolic confinement
\cite{jacak03b}),
 \be \label{formf}
    {\cal F}_{\kvec}^{\mathrm{(e,h)}}
    =e^{-\frac{1}{4}(k_\perp^2 l_{\mathrm{e,h}}^2 +k_z^2 l_z^2)},
 \ee
where $l_{\mathrm{e,h}}$ and $l_z$ are the wavefunction widths in
the dot plane, for electron and hole, and in the growth direction,
respectively, and $k_\perp$ and $k_z$ are the corresponding
components of the phonon wavevector.

Note that the most effectively coupled phonon modes correspond to
wavelengths around the dot size. For gapless bosons with linear
dispersion, their frequency determines the reservoir memory times
and sets up the timescale of non-Markovian effects which, in the
case of the QD system, may be interpreted as ``dressing'' the
localized carriers with coherent lattice deformation field (cf.
\cite{Alicki-spinbozon}). This kind of dynamics has been described
in the limit of very rapid gating pulses
\cite{jacak03b,krummheuer02} and has been shown to lower the
degree of coherent control over the system and to destroy the
coherence of polarization oscillations \cite{krummheuer02,vagov}.
Due to relatively large dot sizes and low sound speed the cut-off
frequency ($\sim 1$ meV) is lower than the transition energy
between different carrier states in a small, self-assembled dot
and the corresponding 1 ps times are experimentally accessible.

For large enough gate duration only the low-frequency sector
contributes and the coupling (\ref{coupling}) may be approximated
by
\begin{equation}
f_{\kvec}
\approx \frac{1}{2}(\sigma_{\mathrm{e}}-\sigma_{\mathrm{h}})
\sqrt{\frac{k}{2\varrho v c}},
\end{equation}
leading, according to (\ref{spectral}), to
\begin{equation}
R(\omega)\simeq R_{0} \omega^{3},\;\;\;
R_{0}= \frac{(\sigma_{\mathrm{e}}-\sigma_{\mathrm{h}})^{2}}%
{16\pi^{2}\varrho c^{5}}.
\end{equation}

Hence, Eqs. (\ref{eq:minerror}) are applicable and one finds for
the specific material parameters of GaAs,
\begin{equation}
\tau_{\mathrm{g}}=\alpha^{2/3} 1.47\; \mathrm{ps},\;\;
\delta_{\min}=\alpha^{2/3} 0.0035.
\end{equation}

The full solution within the proposed model, taking into account
the phonon cut-off for an anisotropic shape ($l_{z}<l_{e,h}$),
according to Eqs. (\ref{formf}), (\ref{coupling}), and allowing
finite temperatures by numerically calculating the spectral
density (\ref{spectral}), is shown in Fig. \ref{fig:result}. The
size-dependent cut-off is reflected by a shift of the optimal
parameters for the two dot sizes: larger dots admit slightly
faster gates and lead to lower error. Interestingly, the trade-off
becomes more apparent at nonzero temperature.

The presence of the upper cut-off could suggest applying the
dynamical decoupling (DD) technique \cite{dd} to diminish
decoherence. However, for high frequencies many other mechanisms
of decoherence become relevant. On the other hand, combining the
bounded-control version of DD \cite{viola03} with the optimization
proposed here might lead to some reduction of the resulting error.
Such techniques might also be useful for eliminating other sources
of noise, not included in our discussion.

Another possibility to reduce the decoherence effect in the QD
system might be to encode qubits in decoherence-free subspaces
\cite{dfs}. However, the phonon wavelength corresponding to the
optimal range of gating times is comparable to the single dot
size, precluding the necessary collective interaction with the
whole QD system. Thus, for decoherence effects which do not
involve real transitions and taking into account the feasible
system geometry and actual nature of phonon coupling, only a small
decrease of the minimal decoherence may be expected.



In conclusion, we have exhibited the trade-off between two
opposite types of decoherence: usual Markovian damping and
dynamically induced non-Markovian decoherence for a realistic
super-ohmic reservoir. To protect the qubit, opposite elementary
strategies are needed: fast and slow gates, respectively. The
minimization of the overall error leads to optimal speed of gate.
We have shown that the trade-off is present in a semiconductor
implementation of quantum information processing. The Markovian
error is caused by recombination, while the non-Markovian one
occurs if the gate operation is not adiabatic with respect to
lattice modes. We have evaluated the minimal error in this case
for a single qubit gate, showing that the two processes indeed
compete. The optimal gating time ($\sim 1$ ps) sets up the limit
beyond which any further gate speed-up is unfavorable. Even at
this optimal point the trade-off gives rise to a significant
error. For two qubit gates involving single-qubit rotations (as
proposed e.g. in \cite{Rossi-SQUID2000}), the present result gives
a rough lower bound for the error, which is of 1-2 orders of
magnitude higher than that admitted by fault-tolerant schemes
\cite{QEC} known so far ($\simeq 10^{-5}$). However, possible
improvements of the latter schemes cannot be excluded. It follows
also that diminishing the responsible constants
$\gamma_{\mathrm{M}}$ (e.g. by elimination of radiative losses
\cite{troiani03}) and $\gamma_{\mathrm{nM}}$ (by optimizing the
system parameters) is the most important task towards
semiconductor implementation of a quantum computer. It is also
important to explore whether the same effect can occur in other
implementations, as well as to what extent the error avoiding
techniques may be helpful here.



We thank Peter Knight and Martin Plenio for useful feedback to the
first version of this paper. L.~J. and P.~M. are grateful to
T.~Kuhn and V.~M.~Axt for discussions. This work was supported by
EC grants EQUIP (IST-1999-11053), RESQ (IST-2002-37559), SQID
(IST-1999-11311), QUPRODIS (IST-2001-38877), and by the Polish
government grant PBZ-MIN-008/P03/2003. P.M. is grateful to the
Humboldt Foundation for support.




\end{document}